\documentclass[preprint]{revtex4}
\usepackage{graphics}
\usepackage{epsfig}
\usepackage{color}

\newcommand{\fig}[1]{Figure (\ref{#1})} 

\begin{document}
\title{Global and Local: Synchronization and Emergence.}
\author{Mogens H.~Jensen and Leo P.~Kadanoff}

\email{mhjensen@nbi.dk, lkadanoff@gmail.com}
\affiliation{Niels Bohr Institute, University of Copenhagen, Blegdamsvej 17, DK-2100,
Copenhagen, Denmark\\
and\\
The James Franck Institute, The University of Chicago, Chicago,
Illinois 60637. USA
}
\date{\today}

\begin{abstract}
When a dynamical system contains several different modes of oscillations it may
behave in a variety of ways: If the modes oscillate at
their own individual frequencies, it exhibits quasiperiodic
behavior; when the modes lock to one another it becomes synchronized, 
or, as a third possibility, complex chaotic behavior may emerge. 
With two modes present, like an internal oscillation coupled 
to a periodic external signal, one
obtains a highly structured phase diagram that exhibits these
possibilities.  In essence, the details are related to the difference
between rational and irrational numbers. Natural system can display 
fragments of this phase diagram,
thereby offering insights into their dynamical mechanisms.
\end{abstract}


\maketitle

\begin{figure}[htbp]
\includegraphics[height=6cm ]{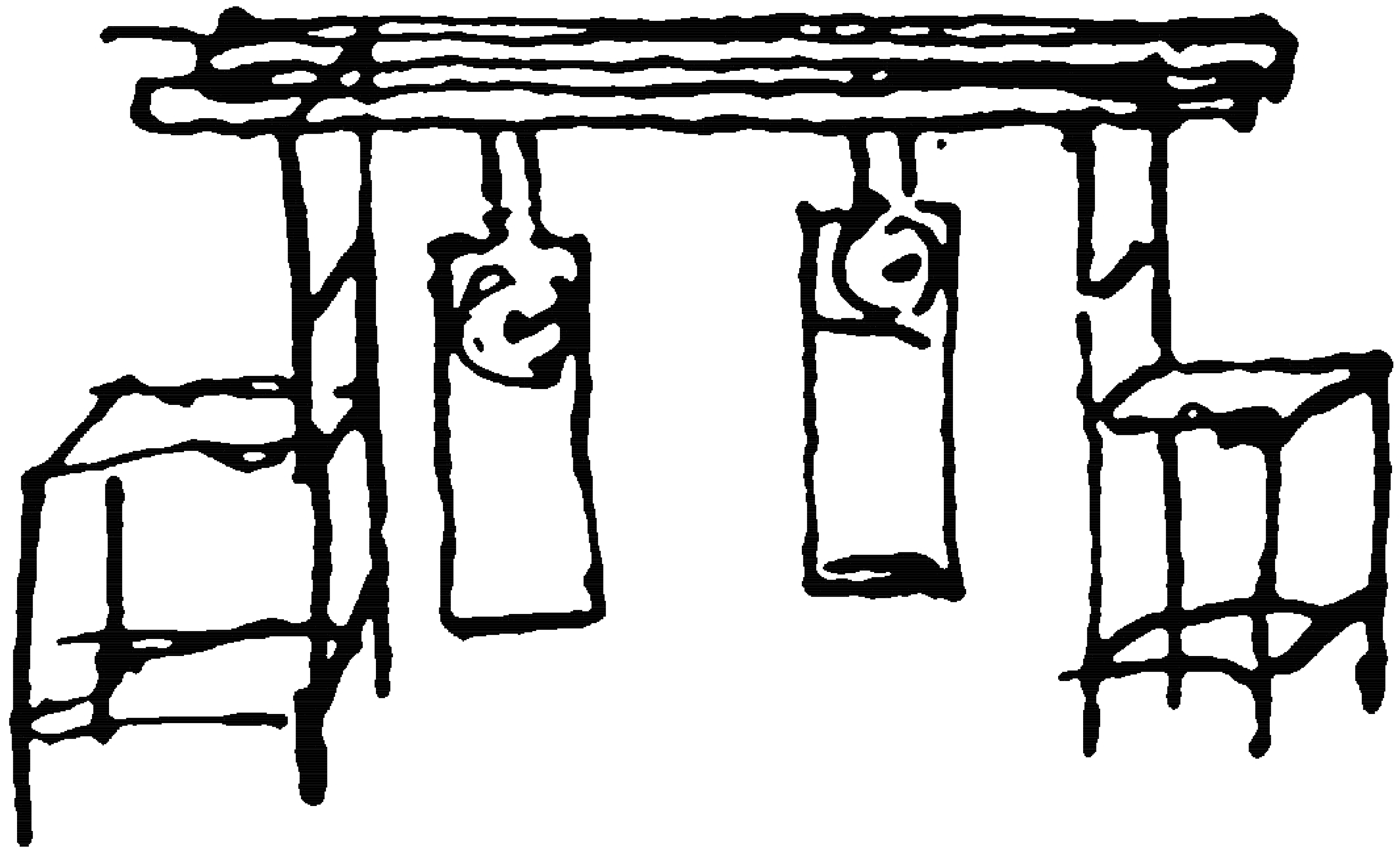}
\caption{Huygens' sketch of his experimental setup for producing
synchronization.  The clocks interact by jiggling the board. Quite soon the
clocks' pendula  lock together their motion  \cite{CH}.
}
\label{Huygens}
\end{figure}

For centuries, physical scientists have studied the synchronization of
oscillators.  In 1665,  Christiaan Huygens  noticed that two clocks hanging
 on a wall
(see \fig{Huygens}) 
 tend to synchronize their pendula \cite{CH}.  A similar scenario occurs
with two metronomes placed on a piano: they interact through vibrations in the
wood and will eventually coordinate their 
motion. In recent years, studies of oscillations and possible synchronizations
\cite{Kurths} have become important
research topics also in the biological sciences.  Living species present us with
a bewildering fauna of oscillators: cell cycles \cite{Ferrell}, 
circadian rhythms \cite{Lefranc}, calcium
oscillations \cite{Goldbeter}, pace maker cells \cite{pacemaker}, protein 
responses \cite{Hoffmann02,Nelson04,Krishna,Mengel},  hormone secretion \cite{hormones}, and so on.
If such oscillators are in the
neighborhood of each other - as they might well be in tissues, organs and cells-
do they tend to synchronize ?

Synchronization presents us with a surprising
depth and complexity, not fully understood to this day.  The phenomenon is at
its basis related to the difference between
rational and irrational numbers!  Each of the two Huygens clocks or the two
metronomes possesses a typical frequency (called an eigenfrequency) $\omega_1$
and
$\omega_2$ respectively. For clocks and metronomes, the two frequencies are most
likely quite similar in magnitude, $\omega_1 \approx \omega_2$ - although maybe
not
exactly equal. When they interact through the wall or through the piano, they
tend to approach each other via a phenomenon called frequency pulling or
frequency locking: they try to synchronize by pulling their frequencies towards
each other such that in the end the two effective frequencies (which we mark by
primes) are exactly equal: $\omega_1' = \omega_2'$.  This corresponds to a
completely synchronized state and
the  more strongly the clocks/metronomes interact the more they tend to
synchronize. Next, let us image that the original (bare) frequencies are such
that $\omega_1 \approx 2 \omega_2$, i.e. one frequency is roughly twice the size
of the
other.  Again, the two oscillators will tend to pull and synchronize at a state
where the effective frequencies are such that $\omega_1' \approx 2 \omega_2'$.  
In general,
a synchronization will be likely as long as $Q \omega_1 \approx P \omega_2$  or
$\omega_1/\omega_2$ can be approximately equal to $P/Q$, where $P$ and $Q$ are
positive 
integer  numbers. In other words, there will tend to be synchronization
when the ratio of the bare frequencies of the two oscillator are close to a
rational number. There exist many rational number, an infinity of them,
but the ones formed from small integers "pull" best. In a synchronized state,
characterized by the rational number, $f= P/Q$, the dynamical motion of the
clocks
is periodic with period  
 $Q$ so that it will return to the its initial state over and over again.

That is one possible behavior.  But it is not the only one.  It is certainly
true that all possible ratios, $\omega_1/\omega_2$ are close to some
$P/Q$ for some integer values of $P$ and $Q$.  However,  as the integers get
larger,
one needs an increasingly close approach between the two ratios  before
synchronization will occur. If there is no rational frequency ratio sufficiently
close,  the system will undergo a motion characterized by an
irrational frequency ratio, f.  The motion is just as orderly as the nearby
periodic orbits, but it never repeats itself.    The required closeness forms a
pattern that depends in  a complex manner upon the integers and upon the
strength of the coupling between the frequencies.

\begin{figure}[htbp]
\center\epsfig{file=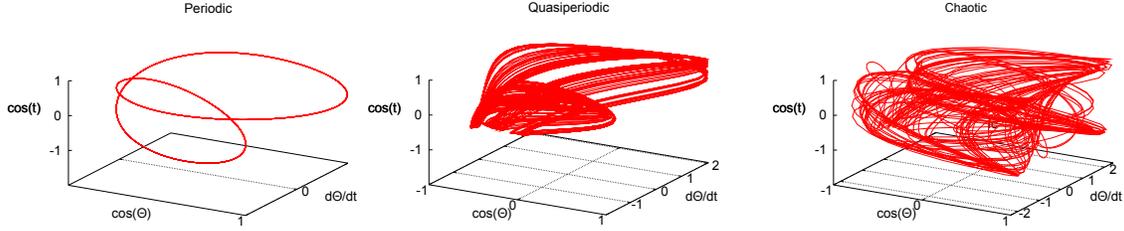,angle=0,width=14.8cm}
\caption{
Three kinds of orbits for two pendula.  In leftmost plate we
see the periodic motion of two locked pendula.  The orbit simply repeats
itself, and forms a one-dimensional curve.  In contrast, the middle plate shows a
quasiperiodic curve formed by two oscillations which retain the independent
frequencies, and in this case, a frequency ratio which is an irrational number. 
Here the motion never repeats itself. It forms a smooth
surface, that is to say it  fills a two dimensional space. 
The rightmost plate presents the result
of a strong coupling in which the motion becomes chaotic.  The orbit is complex
in that the motion is, for some periods, of one type, then switches to another
type of motion with intervals that obey no simple law. The motion never repeats
itself, but it does not fill any simple space. A measurement of the
dimension \cite{GP} of the orbit will show a fractal result, a dimension that is not
equal to one, as in the locked case, nor two as in the quasiperiodic one, but is
instead some number between two and three.
}
\label{orbit}
\end{figure}

\begin{figure}[htbp]
\center\epsfig{file=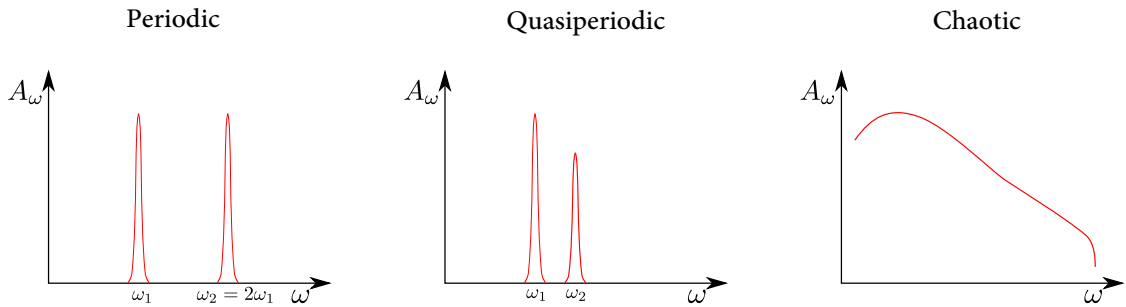,angle=0,width=14.8cm}
\caption{
Schematic plot of frequency spectra for the kinds of orbits shown in
\fig{orbit}.  The three plates correspond to the cases of
periodic/locked,
quasiperiodic, and chaotic motion and show respectively line spectrum with a
fundamental plus overtones, a line spectrum with mixtures of two incommensurate
frequencies, and a smoothly varying spectrum. 
}
\label{fourier}
\end{figure}

Before describing the patterns formed by the different kinds of motion, we
should spend a moment saying how they can be distinguished. The first and
simplest technique is to plot the orbits formed by coordinates of the motion, as
in \fig{orbit}.   As one can see from that figure, there are three kinds of orbits:
locked, quasiperiodic, and chaotic. The orbits look different. They can be
distinguished by their measured dimension \cite{BM,GP,JF} being respectively one, two,
or non-integer in the three cases.  A spectrum generator that measures the
frequency content of each motion will sharply distinguish these patterns. The
periodic motion will have one very sharp frequency line, plus overtones at
multiples of that frequency. The quasiperiodic pattern will show two (or
sometimes more) lines at fundamental frequencies not related by integer
multiplication and then a host of other lines identifiable as integer multiples
of the fundamental frequencies.  In contrast, the chaotic motion will generate a
smoothly varying frequency spectrum as in the third panel of \fig{fourier}.

\begin{figure}[htbp]
\center\epsfig{file=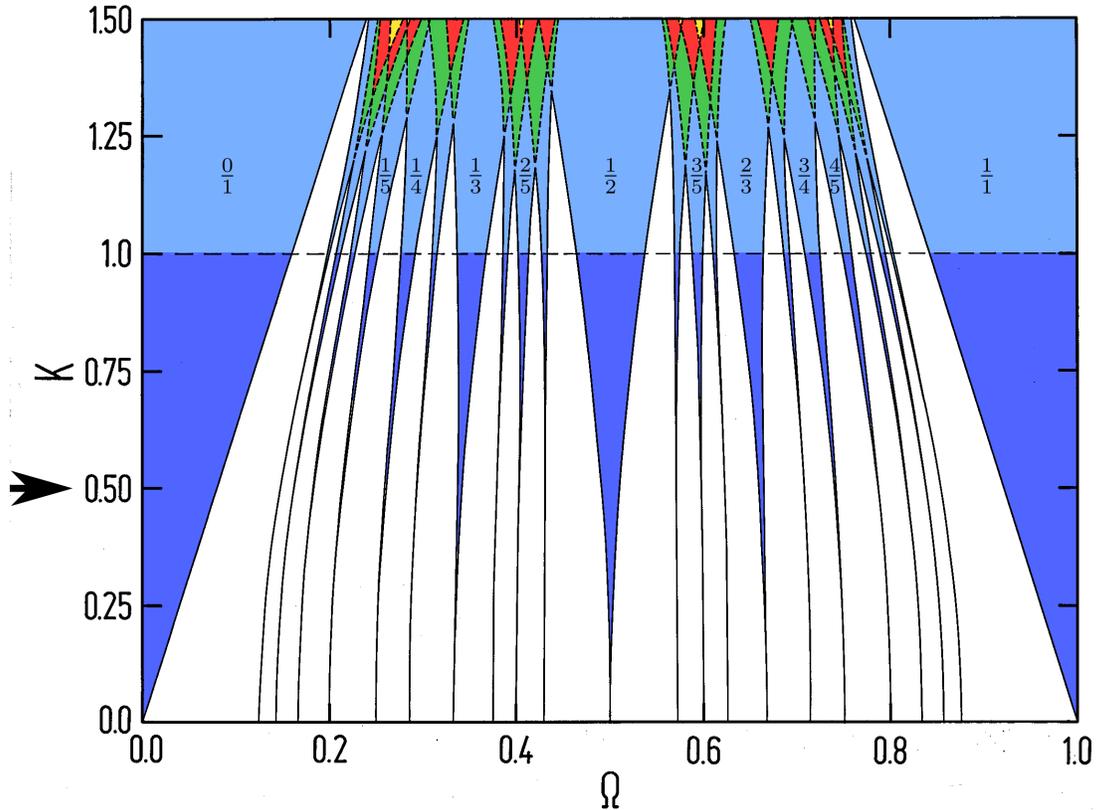,angle=0,width=14.8cm}
\caption{
The pattern of frequency locking and unlocking in a
simplified mathematical model.  The $x$- axis on this graph is the 
original frequency of one of the
two oscillators.  The $y$-axis is the strength of the non-linear interaction
between the oscillators.  The blue regions are ones in which the oscillators
are locked, the numbers attached to each region describes the frequency ratio
for the locking.  The white regions show intermixed quasi-periodic and periodic
behavior, too finely intermingled to be separated by our plot. The green
and red regions
show similarly intermixed behavior, but now also including a chaotic element.
The broken line at  $K=1$ shows the onset of chaotic behavior. Above this line,
chaos is possible (indicated by a change in the blue color); 
below it there is only quasiperiodic and locked behavior.
}
\label{phase}
\end{figure}

It is not just the individual motion that are interesting:  The different kinds
of motions 
 fit together into a richly interwoven phase diagram, see \fig{phase}. 
 This plot shows
how the kinds of motion can depend upon the parameter driving the behavior. 
A.A. Kolmogorov 
has constructed a simplified mathematical model \cite{Kolm}
that can be used  to describe
the pattern of mingling of different kinds of orbits \cite{Quasi,JBB}.
We show that very rich and structured pattern in \fig{phase}. We shall argue
below that elements of this
figure may be found within 
the behavior of frequency
locking in real systems.  But for now, we ask the reader to bear with us and
follow the description of the simplified model.

As mentioned, the tendency to synchronize depends on the interaction strength of
the oscillators. Call this strength $K$  
 - for the clocks it is a measure of the
vibration of the wall, for pace maker cells the intensity of the cell-to-cell
interactions through a  tissue. The results of a simplified model 
is shown in \fig{phase}, and it tells us
how the frequencies can affect the mode locking. On the y-axis is the
strength of the interaction $K$, on the x-axis the bare, original frequency
$\omega_2$ of one of the oscillators.   The other oscillator is assumed to have
bare frequency equal to one. Without any interaction, that is when $K$ is equal
to zero, 
 the oscillators do not interact and the model makes the motion proceed with a
frequency unchanged from the bare value, $\omega_2$. For
irrational values of $\omega_2$ the motion is quasiperiodic, so that
quasiperiodic behavior dominates the $K=0$ line.

As soon as the two oscillators interact just the slightest amount, that is when
$K$
assumes a value just above zero, the model displays frequency pulling so that
it gives a region of
synchronization whenever $\omega_2$ is in  a small interval around each and
every
rational number, $P/Q$. \fig{phase} shows these regions of frequency locking as
blue 
regions.  It has dark blue regions being very narrow at small $K$ and growing wider
as
$K$ gets larger. For each value of the rational number $P/Q$, there will be some
region of locking. These regions are called  called Arnold tongues. They are
named after the  mathematician Vladimir Arnold who did extensive studies of this
model \cite{Arnold}. As $K$ gets larger, the tongues widen and
take up a larger and larger proportion of the frequency interval.

These frequency-locked regions do not exhaust the frequency interval between
zero and
one.  For $K$ between zero and one, 
 in addition to these locked regions, there are a host of quasiperiodic orbits,
motions with
frequency ratios, $f$,  that behave like irrational values of $P/Q$. In these
orbits, the
motion remains orderly but, unlike in the periodic case, it never repeat
itself.  In \fig{phase}, 
the regions of quasiperiodic behavior are shown in white.  These regions also
include infinite numbers of narrow bands of synchronized orbits.

This part of the phase diagram is arranged in a very orderly fashion.  As the
frequency variable on the x-axis increases, each $\omega_2$ gives rise to an
$f$, and these frequency-ratio values all increase with increasing $\omega_2$. 
Here $f$ moves smoothly
through irrational values but gets stuck for an interval, perhaps a very
short interval, on each rational ratio. 


These will be a quasi-periodic orbit for each irrational number. For increased
interaction strength between the oscillators, the synchronization becomes
stronger, and each rational interval tends to widen so that the irrational
'points' become more and more closely packed, maintaining however the property
of having $f$ increase as one moves to the right in the figure. For all $K$
between
zero and one, both kinds of orbits,  synchronized and quasiperiodic, occupy
finite
lengths of the frequency axis.

As $K$ increases further this area covered by the tongues continues to increase,
until at $K=1$, there is only an infinitesimal area left for the irrational
orbits.  At that point these quasiperiodic orbits occupy a "fractal" set
\cite{BM} with a dimension measured to be 0.870 \cite{JBB}.   That the dimension
is less than one is an indication that at
this value of $K$ these orbits cease to occupy a finite fraction of the
frequency axis and instead are relegated to a set of zero length. 
The $K=1$ line thus defines a complementary situation to that at $K=0$: the rational
and irrational numbers have exchanged their roles. Now the rational numbers fill up the
line while the irrationals fill nothing.  At K=1 the irrationals are still all there,  and
they are infinitely more numerous than the rational numbers \cite{HH}, but despite that they
occupy zero length along the line.

At $K=1$, there is a dramatic change in the behavior. The orderly progression
of
quasiperiodic orbits with continuously increasing frequency disappears.
The immediate cause of this change is a region of the flow in which
the derivative of the map defining our model passes through zero.
Just above the $K=1$ line the model begins to  show a
much richer behavior than heretofore.  For some values of the model parameters,
several different orbits, even orbits of different characters, are
simultaneously possible.  Which kind one sees depends upon the initial
conditions of the motion.   Chaotic orbits can be found (see plate c in
\fig{orbit} and \fig{fourier})
 in which the long-term motion is quite unpredictable.  However, regions
of locked motion still exist.  \fig{phase} shows how the locked motion is
intermixed with other kinds of orbits.

Most of the theoretical discussion up to here has been based upon attention to
simplified models.   This attention is justified because real systems, physical,
biological or geological, show many of the same qualitative
features as the models.  The lunar month is incommensurate with the solar year,
and this incommesuration is a real fact about earth. In a women's dorm, women's
"monthly" periods do tend to lock, giving a surprising example of
period-locking \cite{MM} and indicating an unexpected mode of interaction.  It
is said
that locusts pick their 11-, 13-, or 17- year cycle so that other species will
find
it hard to period-lock to them.  We get an insight into the strength of
non-linear couplings by noticing that, for example, predator-prey cycles tend to
be quite chaotic.   In each of these cases, we are analyzing real behavior by
using pictures of simplified behavior, often pictures derived from the study of
simplified models.

Thus, the orbits observed in various sciences  can be referred back to the
orbits of simple dynamical systems.   This referencing to simple models can also
apply to entire regions of phase diagram like that in \fig{phase}.  Each small
region of this diagram is the result of varying parameters in the model over
some small region of parameter space.  Real systems can be expected to show the
same sort of detailed features.  One need only look for these features.  Then by
knowing the region in parameter space, one might hope to get some insights into
biological function or geophysical history.

The point is that some features of models, particularly those involving how
different kinds of motion arise and fit together, are "universal".  That means
that these features are to be found not only in simplified models, but in a wide
variety of circumstances in which the same basic mechanics are at work. 
One such feature is the fractal dimension of the quasiperiodic orbits, 0.87$\dots$,  
at the onset of chaos \cite{JBB}.
Not only does this structure and this dimension appear in the
model, the scenario has been experimentally verified in a number of physical
systems, from onset of
turbulence \cite{Stavans}, Josephson junctions \cite{alstrom, He},
one-dimensional conductors \cite{Gruner}, semiconductors
\cite{Gwinn,Lindsey} and crystals \cite{Martin}.
The number close to 0.87$\dots$ has even been
observed 
in fluids \cite{Stavans}, sliding charge-density-waves \cite{Gruner} and in
Josephson simulators \cite{He}.
Similar studies might see the onset of chaos in biological
systems.  Such onset has been argued to be helpful to biological function
\cite{SK}.
The biological world contains an amazing number of coupled
oscillators.  What does total
synchronization actually mean in a case of cell cycles? And will overlap of
tongues lead to a chaotic state of the cell?  Questions like these can be
asked, and partially answered, by paying attention to model phase diagram like
that of \fig{phase}.


\begin{thebibliography}{2}
\bibitem{Kolm}
The model constructs  successive x-values, $x_0, x_1,..$ from the equation
$x_{k+1} =x_k + \omega_2 +(K/(2\pi)) sin(2 \pi x_k)$.   
The other frequency is given by the period of the sine function and is
$\omega_1=1$.  The  modeler then asks whether the pattern of $x_k$ can be
written as
$x_k=x_0 +f k +G(x_k)$, where $f$ is a rational or irrational frequency ratio,
with
$G$ being a smooth periodic function obeying $G(x+1)=G(x)$. If it can be written
in this manner with a rational $f$, it is said to be periodic; if $f$ is irrational
the result is said to is said to be quasiperiodic.
\end{thebibliography}
\end{document}